\documentclass{PoS}

\def\aj{AJ}                   % Astronomical Journal
\def\apjl{ApJ}                % Astrophysical Journal, Letters
\def\apjs{ApJS}               % Astrophysical Journal, Supplement
\def\aap{A\&A}                % Astronomy and Astrophysics
\def\mnras{MNRAS}             % Monthly Notices of the RAS
\def\nat{Nature}              % Nature
\title{The dependence of optical polarisation of blazars on the synchrotron component peak frequency}
\ShortTitle{The dependence of the blazars on the synchrotron peak frequency}

\author{\speaker{Emmanouil Angelakis}\thanks{Presented on behalf of the entire {\em RoboPol} collaboration (see note at the end of the manuscript)}\\
         Max-Planck-Institut f\"ur Radioastronomie, Auf dem H\"ugel 69, Bonn 53121, Germany \\
        E-mail: \email{eangelakis@mpifr.de}}

\author{Dmitry Blinov\\
        Department of Physics and Institute for Plasma Physics, University of Crete, 71003, Heraklion, Greece\\
        Foundation for Research and Technology - Hellas, IESL, Voutes, 71110 Heraklion, Greece\\
        Astronomical Institute, St. Petersburg State University,Universitetsky pr. 28, Petrodvoretz, 198504 St. Petersburg, Russia\\
        E-mail: \email{blinov@physics.uoc.gr}}

\author{Markus B\"ottcher\\
  North-West University, Potchefstroom Campus, Private Bag X6001, Potchefstroom 2520, South Africa\\
        E-mail: \email{Markus.Bottcher@nwu.ac.za}}

\author{Talvikki Hovatta\\
        Aalto University Mets\"ahovi Radio Observatory, Mets\"ahovintie 114, 02540 Kylm\"al\"a, Finland\\
        E-mail: \email{talvikki.hovatta@aalto.fi}}

\author{Sebastian Kiehlmann\\
        Cahill Center for Astronomy and Astrophysics, California Institute of Technology, 1200 E California Blvd, MC 249-17, Pasadena, CA 91125, USA\\
        E-mail: \email{skiehl@caltech.edu}}

\author{Ioannis Myserlis\\
         Max-Planck-Institut f\"ur Radioastronomie, Auf dem H\"ugel 69, Bonn 53121, Germany \\
        E-mail: \email{imyserlis@mpifr-bonn.mpg.de}}

\author{Vassiliki Pavlidou\\
        Department of Physics and Institute for Plasma Physics, University of Crete, 71003, Heraklion, Greece\\
        Foundation for Research and Technology - Hellas, IESL, Voutes, 71110 Heraklion, Greece\\
        E-mail: \email{pavlidou@physics.uoc.gr}}

\author{J. Anton Zensus\\
         Max-Planck-Institut f\"ur Radioastronomie, Auf dem H\"ugel 69, Bonn 53121, Germany \\
        E-mail: \email{azensus@mpifr-bonn.mpg.de}}

\abstract{The \textit{RoboPol} instrument and the relevant program was developed in order to conduct a systematic study of the optical polarisation variability of blazars. Driven by the discovery that long smooth rotations of the optical polarisation plane can be associated with the activity in other bands and especially in gamma rays, the program was meant to investigate the physical mechanisms causing them and quantify the optical polarisation behaviour in blazars. Over the first three nominal observing seasons (2013, 2014 and 2015) \textit{RoboPol} detected 40 rotations in 24 blazars by observing a gamma--ray-loud and gamma--ray-quite unbiassed sample of blazars, providing a reliable set of events for exploring the phenomenon. The obtain datasets provided the ground for a systematic quantification of the variability of the optical polarisation in such systems. In the following after a brief review of the discoveries that relate to the gamma-ray loudness of the sources we move on to discuss a simple jet model that explains the observed dichotomy in terms of polarisation between gamma--ray-loud and quite sources and the dependence of polarisation and the stability of the polarisation angle on the synchrotron peak frequency.}

\FullConference{7th Fermi Symposium 2017\\
		15-20 October 2017\\
		Garmisch-Partenkirchen, Germany}

\begin{document}

\section{Introduction}

The first detection of a smooth and long rotation of the optical polarisation plane i.e. the electric vector position angle (EVPA), was reportedly found in OJ287 \cite{1988A&A...190L...8K}. The attention of the community was however intrigued when \cite{2008Natur.452..966M} reported a similar event in BL Lacertae that was associated with a radio-to-gamma-ray outburst. On the basis of their observation that "a bright feature in the jet" induced "a double flare" observed from optical bands to TeV energies, as well as "a delayed outburst at radio wavelengths", they concluded that the event started in a region with a helical magnetic field configuration. They identified that region "with the acceleration and collimation zone predicted by the theories". The latest brightening of the feature was attributed to its crossing "a standing shock wave" which appeared as the bright radio "core". The importance of the discovery lies mostly on the fact that it offers a way to probe the inner jet of Active Galactic Nuclei in the region where the the jet gets accelerated and collimated in an an environment of coiled magnetic field. Thus, address the very question of the jet formation. Later studies (e.g. \cite{2010Natur.463..919A}, \cite{2010ApJ...710L.126M} etc.) proposed several alternativve explanations; all however inevitably based on hand-picked events given the transit character of the phenomenon. The \textit{RoboPol} program \cite{2014MNRAS.442.1706K}, \cite{2014MNRAS.442.1693P} was developed to conduct a systematic study of the phenomenon alongside with quantifying the optical polarisation of blazars and its variability.

Starting from the Second \textit{Fermi} Source Catalog (2FGL) \cite{2012ApJS..199...31N} and after excluding $10^\circ$ on either side of the galactic plane, setting a minimum integrated photon flux at around $10^{-8}$~cm$^{-2}$~s$^{-1}$ and complying with observational constrains, we composed a sample of 62 gamma--ray-loud (GL) blazars to be monitored (for details c.f. \cite{2014MNRAS.442.1693P}). A comparison gamma--ray-quite (GQ) sample was drawn from the the 15~GHz OVRO monitoring sample \cite{2014MNRAS.438.3058R}. The selection was based on a the amplitude of radio variability ($> 5~\%$), the radio mean flux density (at least 60~mJy), the absence from the 2FGL, and again the observability from Skinakas telescope.    
 
In the following we start with a review of angle domain findings that relate with the gamma-ray activity of the sources. Later we focus our attention on the amplitude domain and examine the dichotomy between GL and GQ sources. Finally we propose a jet model that can explain the observations and especially the dependence of both the degree of polarisation and the uniformity of the angle on the synchrotron peak frequency.    

The convention we adopt for the definition of a rotations includes: a continuous EVPA change larger than $90^\circ$, comprised by at least four consecutive measurements with significant swings
between them.

\section{EVPA rotations and gamma-ray activity}

The first three seasons of nominal \textit{RoboPol} operation (2013, 2014 and 2015) revealed a total of 40 EVPA rotations that happened in 24 blazars summarised in \cite{2015MNRAS.453.1669B} and \cite{2016MNRAS.457.2252B}. Before robopol a mere total of 16 events had been found in 10 sources. As it is discussed there the EVPA rotations do not seem to be favouring any of the source classes that were examined. High and Low-Synchrotron Peaked (LSP, HSP) sources can show rotations alike as much as BL Lacs and quasars or TeV and non-TeV emitting ones. Interestingly, the rotation rate (deg~day$^{-1}$) can vary noticeably for the same source from one event to the other and so is the sense of rotation. Given the adopted conditions needed for a sequence of measurements to constitute a rotation, all detected events are in GL sources. This already indicates a relevance with the gamma-ray loudness. Within the uncertainties, there is no indication that for an "orphan" rotation which is unassociated with some activity in gamma rays. There has not been any rotation that occurred with a non-zero lag from the associated gamma-ray event (Blinov et al. submitted). Finally, Figure~8 of \cite{2016MNRAS.462.1775B} shows that the "rotators" are both more luminous and more variable in gamma-rays.  

\section{The polarisation of GL and GQ sources}

Already on the basis of single-shot measurement obtained during the \textit{RoboPol} instrument commissioning, \cite{2014MNRAS.442.1693P} noted that GL sources as a population are significantly more polarised that GQ ones. Using the datasets collected during the 2013 and 2014 observing seasons and after implementing a maximum likelihood analysis that accounts for uneven sampling and measurement uncertainties, \cite{2016MNRAS.463.3365A} confirmed beyond a $4\sigma$ level that GL sources are on average more polarised ($\sim9~\%$) than GQ ones ($\sim3~\%$).

\section{The jet model}

A jet populated by a helical magnetic field and possibly an underlying turbulent one, was though to be adequate for explaining the observed dichotomy. Occasional mild shocks would locally energise the particles and increase the local field uniformity. The particles in the small volume immediately downstream the shock will emit the high energy photons. Since this emission originates at an environment of increased field uniformity it will be associated with higher polarisation. As the particles propagate downstream loosing energy, the associated low energy photons are integrated over larger volumes with different orientations of the field.This naturally being associated with lower polarisation. Hence, high energy emission should be more polarised that low energy one.    

In this context, GL which are know to be jet dominated and intensely variable, show often impulsive events of particle acceleration. They hence often reach the necessary high energy limit to produce high energy and high polarisation emission as well as reach gamma-ray detection limits and be classified as GL ones. On the contrary, GQ sources are deprived of such activity. The rarely managed to produce high energy photons and hence are characterised by lower polarisation. In parallel they barely ever pass the detection threshold that would classify them as GL sources.   

Though simple, this model includes all the basic components of relativistic jets as they are thought today. Beyond the explanation it offers for the dichotomy of the GL and GQ sources, it also provides a natural explanation of the observed dependence of the polarisation on the synchrotron component peak frequency as we discuss in the next. 

\section{The polarisation dependence on the synchrotron peak frequency}

As we show in Figure~5 of \cite{2016MNRAS.463.3365A}, the degree of optical polarisation drops with the frequency of the peak in the synchrotron SED component. So that HSP sources deliver less polarised optical emission compared to sources with their peak of the synchrotron component at much lower energies (LSP). For the LSP sources the synchrotron peak appears in the vicinity of IR so that the optical emission observed by \textit{RoboPol} constitutes the high energy part of their spectrum. So as we said earlier high energy photons must be associated with higher polarisation. For the HSP sources on the other hand the optical emission corresponds to low energy part of their SED as their synchrotron peak lives in UV or even higher bands. Hence the low energy emission will be characterised by lower polarisation.  
   
This picture further implies a predictable behaviour for the EVPA whose "randomness" should depend also on the synchrotron peak frequency. The topic is discussed immediately.    

\section{The dependence of the EVPA randomness on the synchrotron peak frequency}

\begin{figure}[] 
\centering
 \includegraphics[trim={0 0 0 0},clip, width=.95\textwidth]{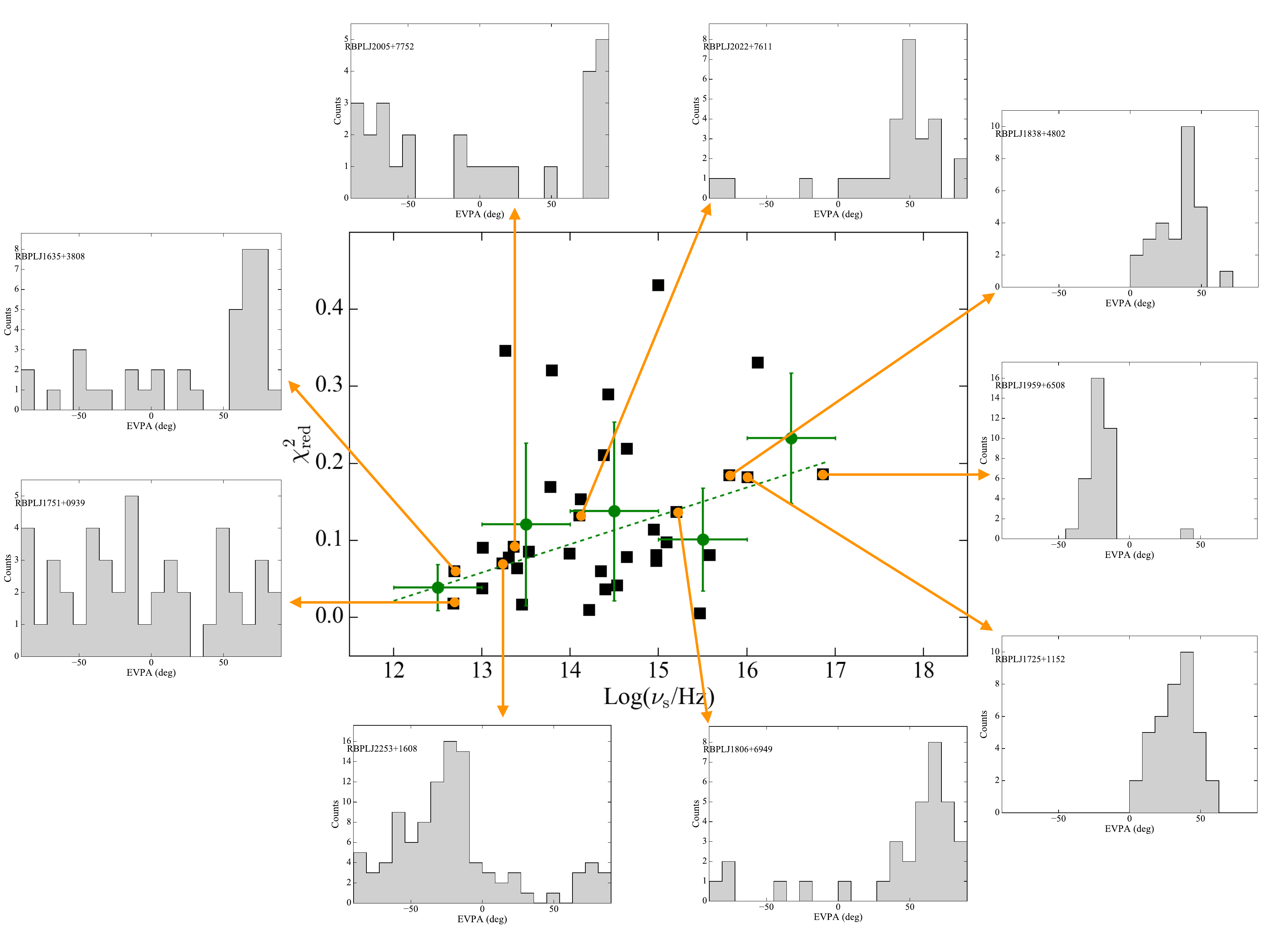} 
\caption{The increase of the EVPA stability as a function of the synchrotron peak frequency. The EVPA stability is measured with the reduced $\chi^2$ of the comparison with a uniform distribution. The satellite panels show the PDFs for some example cases of progressively (from left to right) increasing EVPA stability. }
\label{fig:evpa_synch}
\end{figure}

As we discussed earlier the optical emission from HSP sources should be coming roughly for a region much larger than that from the LSP ones. The presence of turbulent field component in a larger region would not be significant and the helical component would be dominant. Further more, any variability would be happening slower that for a smaller region which would  include a smaller number of random field cells. Statistically then, the EVPA of HSPs would tend to align with a preferred direction. That is, a plot of EVPA "stability" would increase with increasing synchrotron peak frequency. Figure~6 in \cite{2016MNRAS.463.3365A} shows that this prediction is indeed seen. Although the data points there do not show a correlation of high significance, the fitted slope is so. For some example cases we show the gradual increase if the EVPA stability with the synchrotron peak position in Figure~\ref{fig:evpa_synch}. In the plot only cases with a minimum of 20 measurements are shown. As it becomes apparent, towards higher frequencies the EVPA tends to concentrate around a preferred direction thus its PDF appears less uniform (or equivalently the reduced $\chi^2$ increases).      
% -----------------------------------------------------------------------
% -----------------------------------------------------------------------

\section{The orientation of the non-uniform EVPAs}

According to our model which predicts the dominance of the helical over the turbulent field component in HSP sources and given the small angles to the line of site for blazars, one would expect that the projected magnetic field would orient itself perpendicularly to the jet axis. As a consequence, the EVPA -- a proxy of the projected magnetic field -- would appear in close alignment to the jet axis in HSPs as the optical emission is in the optically thin part of the synchrotron component. 

\begin{figure}[] 
\centering
\begin{tabular}{ccc}
 \includegraphics[trim={100 10 170 0},clip, width=.31\textwidth]{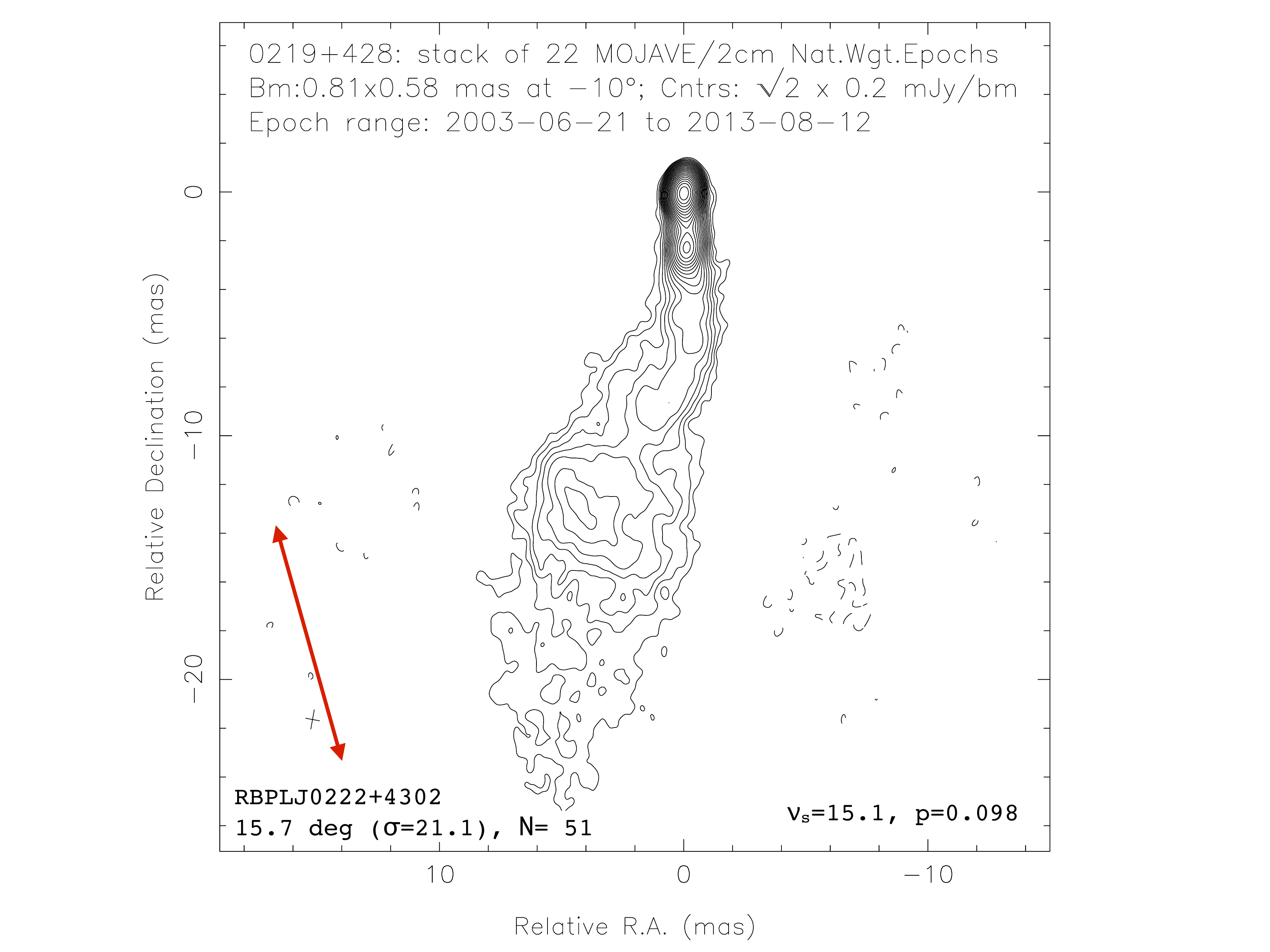} & \includegraphics[trim={100 10 170 0},clip, width=.31\textwidth]{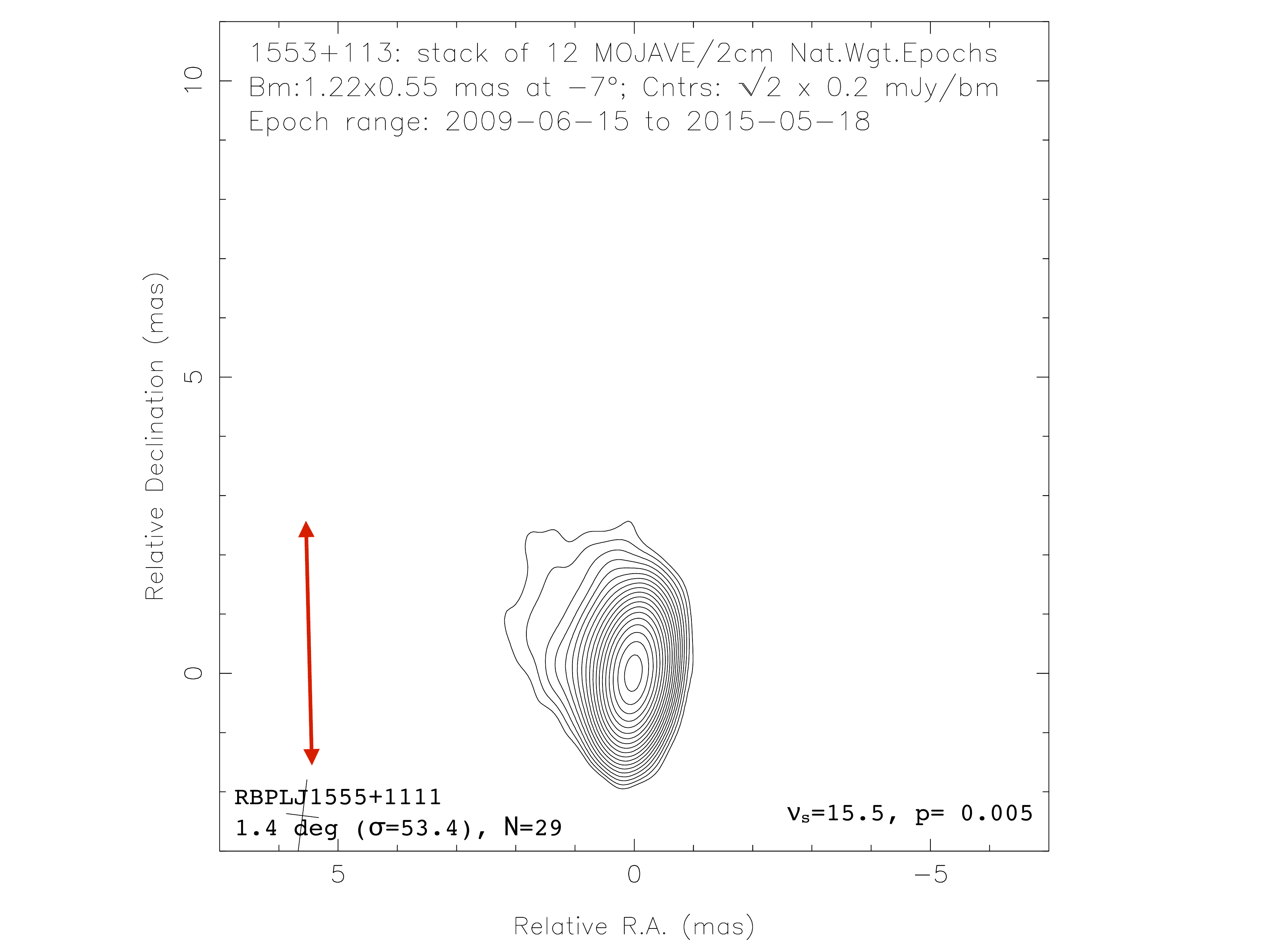} &\includegraphics[trim={100 10 170 0},clip, width=.31\textwidth]{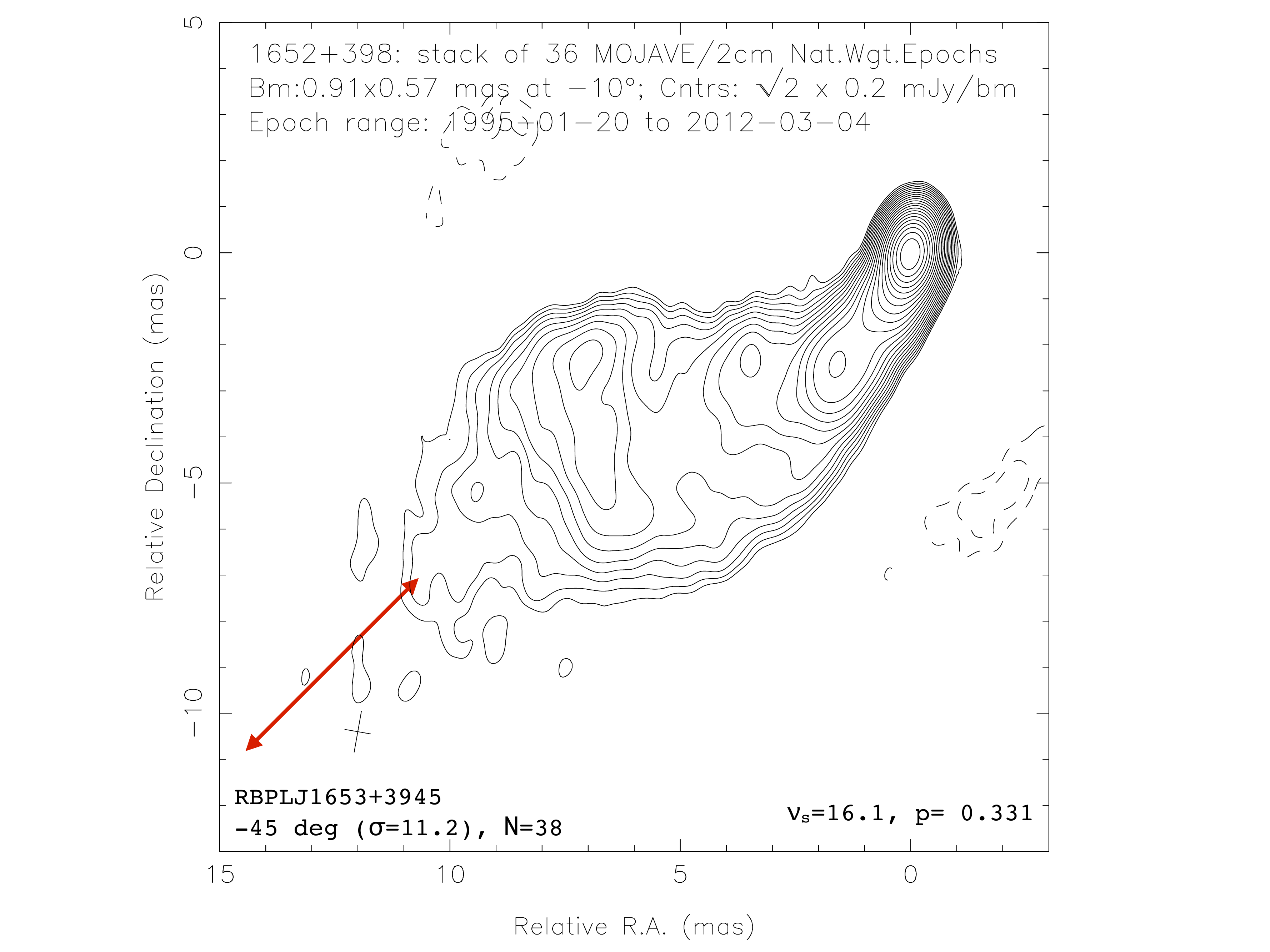} \\
  \includegraphics[trim={100 10 170 0},clip, width=.31\textwidth]{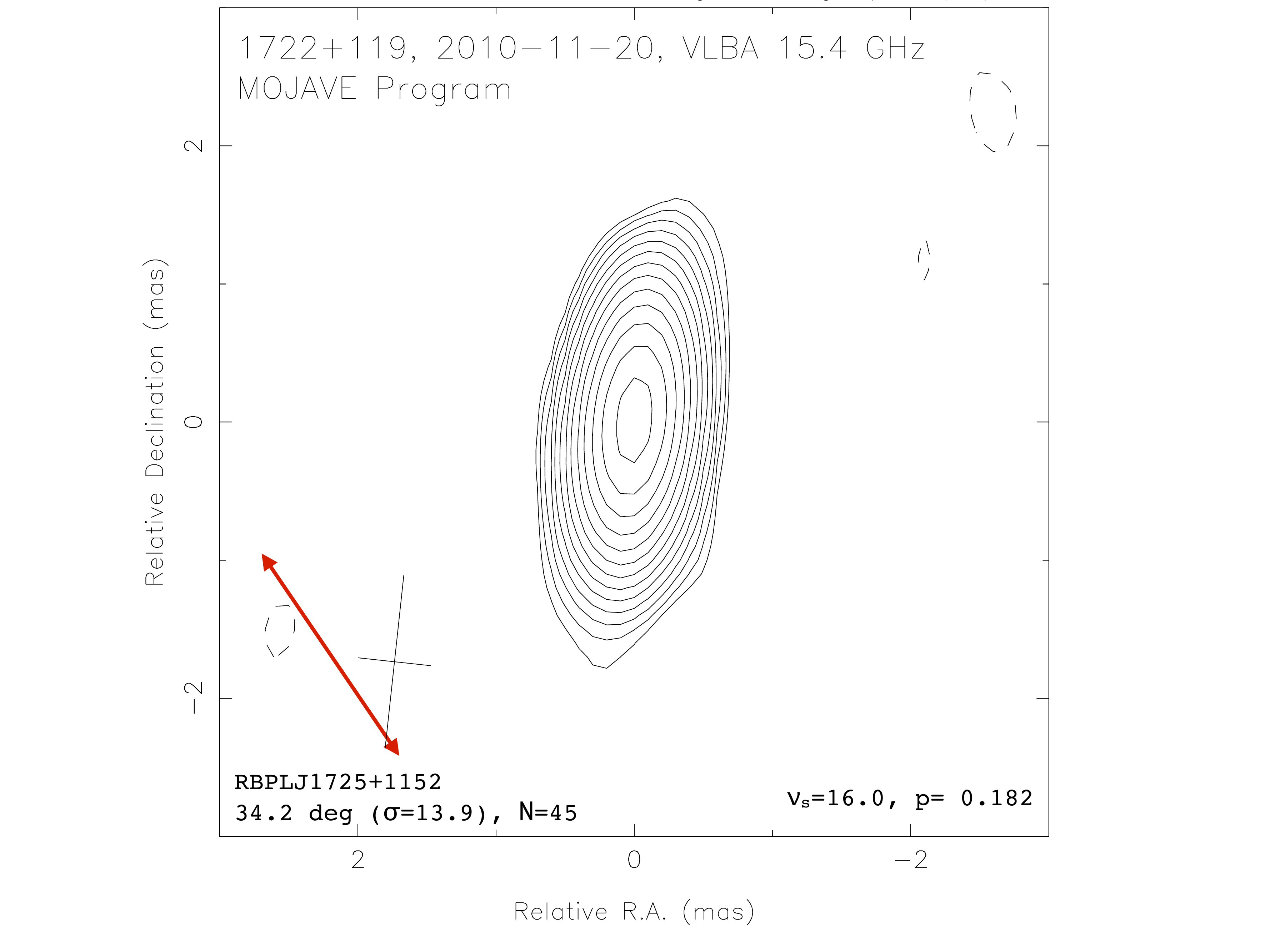} &\includegraphics[trim={100 10 170 0},clip, width=.31\textwidth]{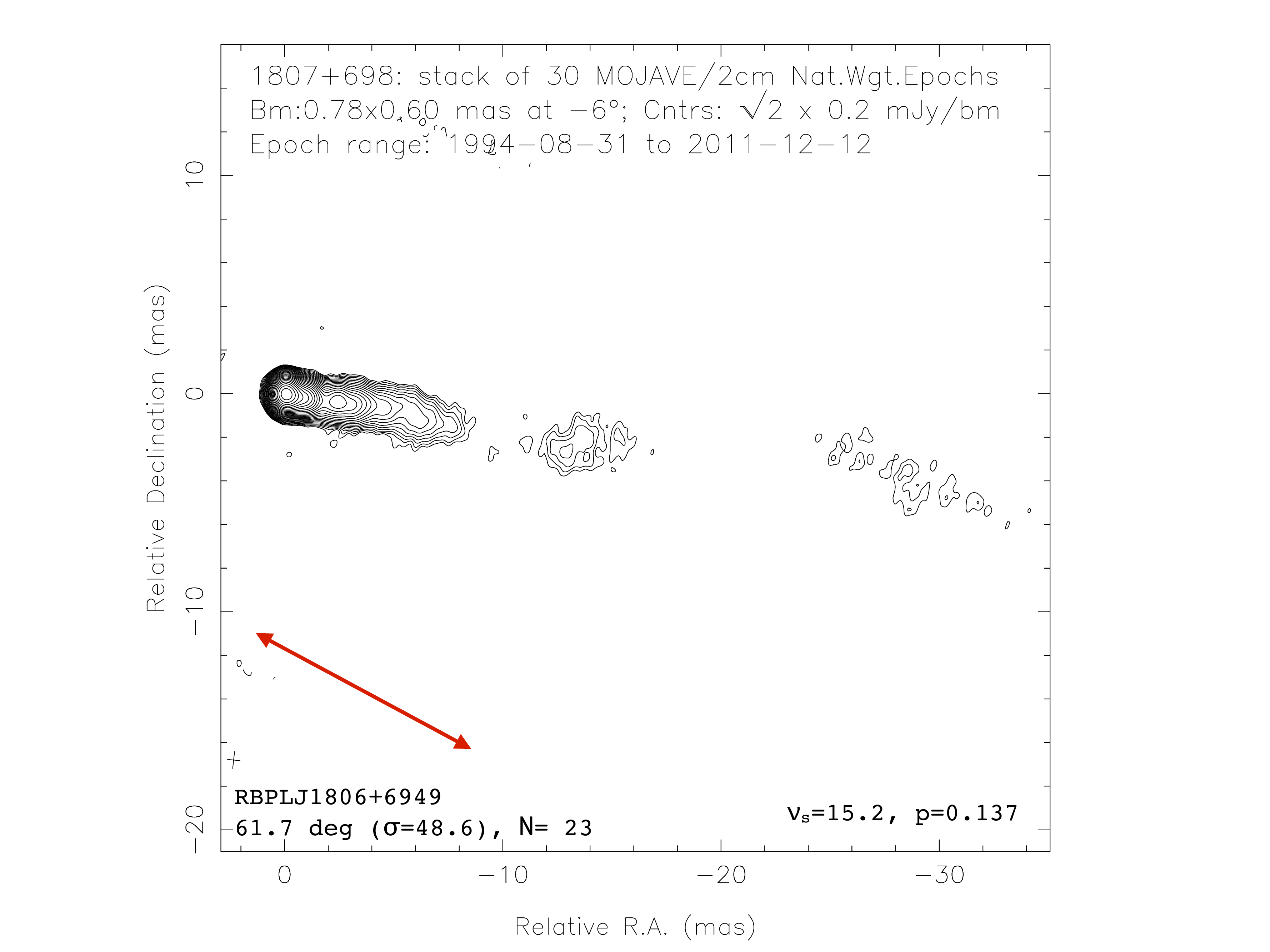} & \includegraphics[trim={100 10 170 0},clip, width=.31\textwidth]{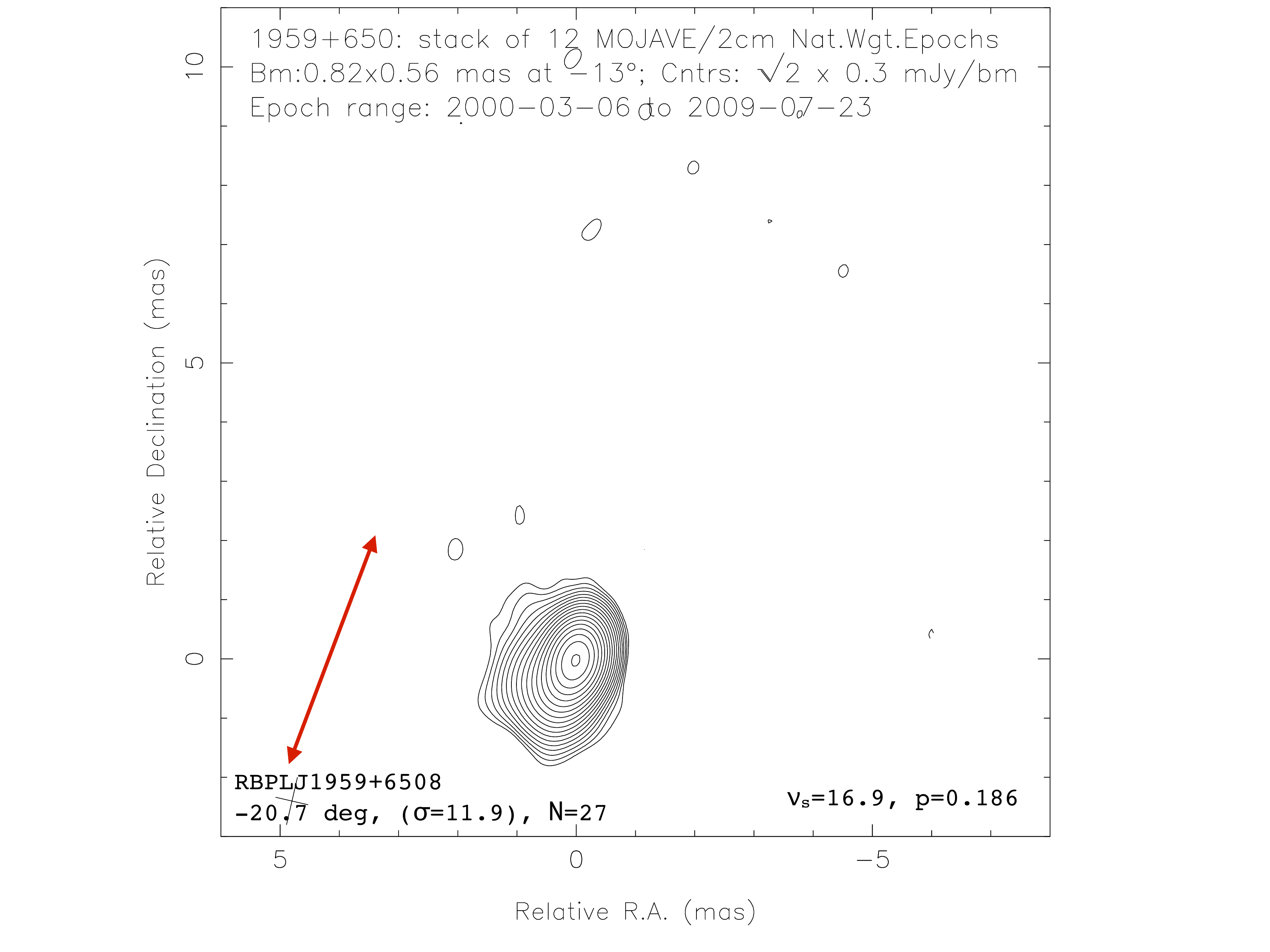} \\
 \includegraphics[trim={100 10 170 0},clip, width=.31\textwidth]{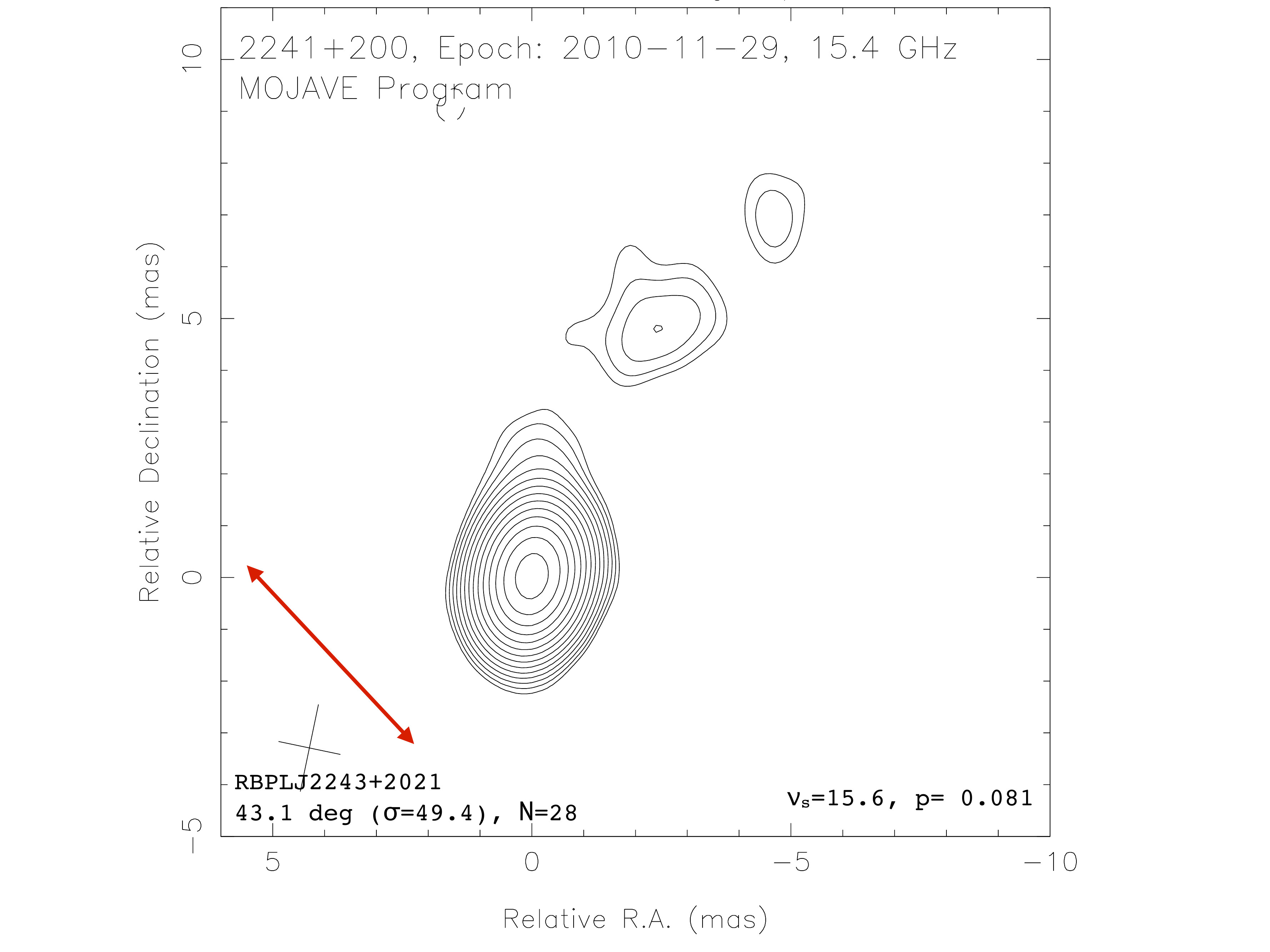} &  \\
\end{tabular}
\caption{The 15 GHz MOJAVE VLBA images \cite{2009AJ....137.3718L} of the sources with $\log\left({\nu_\mathrm{s}/\mathrm{Hz}}\right)\ge 15$. For those our model forecasts a close alignment of the EVPA and the radio jet. Except for the case of 2241$+$200 the maps have been subjected to natural weighting. The red double headed arrow denotes the median EVPA. In the lower part of each plot we also give the median EVPA, its sigma and the number of measurements (lower left), along with the logarithm of the synchrotron peak frequency and the median polarisation fraction (lower right).}
\label{fig:maps_hsp}
\end{figure}

The best study of this prediction came from \cite{2016A&A...596A..78H} (Figures~7 and 8) who conducted  careful study of the optical polarisation of high energy BL Lacertae objects in order to identify possible systematic differences with sources that appear in the TeV sky. for a limited sample of sources for which the EVPA as well as the jet orientation was known, they found that indeed six out of nine sources orient their EVPA within 20 degrees from the jet axis implying that the projected magnetic field is indeed perpendicular to the jet. This is the best test of our hypothesis so far as these sources would populate the right-hand part of Figure~\ref{fig:evpa_synch}. Unfortunately, the lack of high angular resolution radio observations will make the expansion of this sample challenging. For all the sources in Figure~\ref{fig:evpa_synch} with synchrotron peak frequency above $10^{15}$~Hz that are in the MOJAVE sample \cite{2009AJ....137.3718L}, we show the orientation of the EVPA with respect to the jet in Figure~\ref{fig:maps_hsp}. For the cases with a clear jet presence a preference towards close alignement of the EVPA to the jet axis is also observed.  
% -----------------------------------------------------------------------
% -----------------------------------------------------------------------

\section*{Acknowledgments}

The {\em RoboPol} program is a collaboration between: {\it University of Crete (Heraklion, Greece)}, {\it
  Max-Planck-Institut f\"ur Radioastronomie (Germany)}, {\it California Institute of Technology (USA)}, {\it
  Inter-University Centre for Astronomy and Astrophysics (India)} and the {\it Torun Centre for Astronomy,
  Nicolaus Copernicus University (Poland) }. This research has made use of data from the MOJAVE database that is maintained by the MOJAVE team \cite{2009AJ....137.3718L}. We wish to thank the internal MPIfR referee Dr C. Casadio for the careful reading and the thorough comments.

\bibliographystyle{JHEP} % style aa.bst
%\bibliography{/Users/mangel/work/Literature/MyBIB/References.bib} % your references Yourfile.bib

\providecommand{\href}[2]{#2}\begingroup\raggedright\begin{thebibliography}{10}

\bibitem{1988A&A...190L...8K}
S.~{Kikuchi}, Y.~{Mikami}, M.~{Inoue}, H.~{Tabara} and T.~{Kato}, \emph{{A
  synchronous variation of polarization angle in OJ 287 in the optical and
  radio regions}}, {\emph{\aap} {\bf 190} (Jan., 1988) L8--L10}.

\bibitem{2008Natur.452..966M}
A.~P. {Marscher}, S.~G. {Jorstad}, F.~D. {D'Arcangelo}, P.~S. {Smith}, G.~G.
  {Williams}, V.~M. {Larionov} et~al., \emph{{The inner jet of an active
  galactic nucleus as revealed by a radio-to-{$\gamma$}-ray outburst}},
  \href{http://dx.doi.org/10.1038/nature06895}{\emph{\nat} {\bf 452} (Apr.,
  2008) 966--969}.

\bibitem{2010Natur.463..919A}
A.~A. {Abdo}, M.~{Ackermann}, M.~{Ajello}, M.~{Axelsson}, L.~{Baldini},
  J.~{Ballet} et~al., \emph{{A change in the optical polarization associated
  with a {$\gamma$}-ray flare in the blazar 3C279}},
  \href{http://dx.doi.org/10.1038/nature08841}{\emph{\nat} {\bf 463} (Feb.,
  2010) 919--923}, [\href{https://arxiv.org/abs/1004.3828}{{\tt 1004.3828}}].

\bibitem{2010ApJ...710L.126M}
A.~P. {Marscher}, S.~G. {Jorstad}, V.~M. {Larionov}, M.~F. {Aller}, H.~D.
  {Aller}, A.~{L{\"a}hteenm{\"a}ki} et~al., \emph{{Probing the Inner Jet of the
  Quasar PKS 1510-089 with Multi-Waveband Monitoring During Strong Gamma-Ray
  Activity}},
  \href{http://dx.doi.org/10.1088/2041-8205/710/2/L126}{\emph{\apjl} {\bf 710}
  (Feb., 2010) L126--L131}, [\href{https://arxiv.org/abs/1001.2574}{{\tt
  1001.2574}}].

\bibitem{2014MNRAS.442.1706K}
O.~G. {King}, D.~{Blinov}, A.~N. {Ramaprakash}, I.~{Myserlis}, E.~{Angelakis},
  M.~{Balokovi{\'c}} et~al., \emph{{The RoboPol pipeline and control system}},
  \href{http://dx.doi.org/10.1093/mnras/stu176}{\emph{\mnras} {\bf 442} (Aug.,
  2014) 1706--1717}, [\href{https://arxiv.org/abs/1310.7555}{{\tt 1310.7555}}].

\bibitem{2014MNRAS.442.1693P}
V.~{Pavlidou}, E.~{Angelakis}, I.~{Myserlis}, D.~{Blinov}, O.~G. {King},
  I.~{Papadakis} et~al., \emph{{The RoboPol optical polarization survey of
  gamma-ray-loud blazars}},
  \href{http://dx.doi.org/10.1093/mnras/stu904}{\emph{\mnras} {\bf 442} (Aug.,
  2014) 1693--1705}, [\href{https://arxiv.org/abs/1311.3304}{{\tt 1311.3304}}].

\bibitem{2012ApJS..199...31N}
P.~L. {Nolan}, A.~A. {Abdo}, M.~{Ackermann}, M.~{Ajello}, A.~{Allafort},
  E.~{Antolini} et~al., \emph{{Fermi Large Area Telescope Second Source
  Catalog}}, \href{http://dx.doi.org/10.1088/0067-0049/199/2/31}{\emph{\apjs}
  {\bf 199} (Apr., 2012) 31}, [\href{https://arxiv.org/abs/1108.1435}{{\tt
  1108.1435}}].

\bibitem{2014MNRAS.438.3058R}
J.~L. {Richards}, T.~{Hovatta}, W.~{Max-Moerbeck}, V.~{Pavlidou}, T.~J.
  {Pearson} and A.~C.~S. {Readhead}, \emph{{Connecting radio variability to the
  characteristics of gamma-ray blazars}},
  \href{http://dx.doi.org/10.1093/mnras/stt2412}{\emph{\mnras} {\bf 438} (Mar.,
  2014) 3058--3069}, [\href{https://arxiv.org/abs/1312.3634}{{\tt 1312.3634}}].

\bibitem{2015MNRAS.453.1669B}
D.~{Blinov}, V.~{Pavlidou}, I.~{Papadakis}, S.~{Kiehlmann}, G.~{Panopoulou},
  I.~{Liodakis} et~al., \emph{{RoboPol: first season rotations of optical
  polarization plane in blazars}},
  \href{http://dx.doi.org/10.1093/mnras/stv1723}{\emph{\mnras} {\bf 453} (Oct.,
  2015) 1669--1683}, [\href{https://arxiv.org/abs/1505.07467}{{\tt
  1505.07467}}].

\bibitem{2016MNRAS.457.2252B}
D.~{Blinov}, V.~{Pavlidou}, I.~E. {Papadakis}, T.~{Hovatta}, T.~J. {Pearson},
  I.~{Liodakis} et~al., \emph{{RoboPol: optical polarization-plane rotations
  and flaring activity in blazars}},
  \href{http://dx.doi.org/10.1093/mnras/stw158}{\emph{\mnras} {\bf 457} (Apr.,
  2016) 2252--2262}, [\href{https://arxiv.org/abs/1601.03392}{{\tt
  1601.03392}}].

\bibitem{2016MNRAS.462.1775B}
D.~{Blinov}, V.~{Pavlidou}, I.~{Papadakis}, S.~{Kiehlmann}, I.~{Liodakis},
  G.~V. {Panopoulou} et~al., \emph{{RoboPol: do optical polarization rotations
  occur in all blazars?}},
  \href{http://dx.doi.org/10.1093/mnras/stw1732}{\emph{\mnras} {\bf 462} (Oct.,
  2016) 1775--1785}, [\href{https://arxiv.org/abs/1607.04292}{{\tt
  1607.04292}}].

\bibitem{2016MNRAS.463.3365A}
E.~{Angelakis}, T.~{Hovatta}, D.~{Blinov}, V.~{Pavlidou}, S.~{Kiehlmann},
  I.~{Myserlis} et~al., \emph{{RoboPol: the optical polarization of
  gamma-ray-loud and gamma-ray-quiet blazars}},
  \href{http://dx.doi.org/10.1093/mnras/stw2217}{\emph{\mnras} {\bf 463} (Dec.,
  2016) 3365--3380}, [\href{https://arxiv.org/abs/1609.00640}{{\tt
  1609.00640}}].

\bibitem{2016A&A...596A..78H}
T.~{Hovatta}, E.~{Lindfors}, D.~{Blinov}, V.~{Pavlidou}, K.~{Nilsson},
  S.~{Kiehlmann} et~al., \emph{{Optical polarization of high-energy BL Lacertae
  objects}}, \href{http://dx.doi.org/10.1051/0004-6361/201628974}{\emph{\aap}
  {\bf 596} (Dec., 2016) A78}, [\href{https://arxiv.org/abs/1608.08440}{{\tt
  1608.08440}}].

\bibitem{2009AJ....137.3718L}
M.~L. {Lister}, H.~D. {Aller}, M.~F. {Aller}, M.~H. {Cohen}, D.~C. {Homan},
  M.~{Kadler} et~al., \emph{{MOJAVE: Monitoring of Jets in Active Galactic
  Nuclei with VLBA Experiments. V. Multi-Epoch VLBA Images}},
  \href{http://dx.doi.org/10.1088/0004-6256/137/3/3718}{\emph{\aj} {\bf 137}
  (Mar., 2009) 3718--3729}, [\href{https://arxiv.org/abs/0812.3947}{{\tt
  0812.3947}}].

\end{thebibliography}\endgroup

%\begin{thebibliography}{99}
%\bibitem{...}
%....%
\providecommand{\href}[2]{#2}\begingroup\raggedright\endgroup

%\end{thebibliography}

\end{document}